# Charge emissions from Electrosprays in vacuum: mixtures of formamide with methylammonium formate

David Garoz,[1, 2] and Juan Fernández de la Mora[2]

[1]ETSI Aeronáuticos, Universidad Politécnica de Madrid, Madrid, 28040, Spain

[2]Department of Mechanical Engineering and Materials Science, Yale University, New Haven, CT 06520-8267, USA

The charge/mass distribution f($q/m$) of nanodrops and ions electrosprayed in vacuum from mixtures of formamide (FM) and methylammonium formate (MAF) is studied by time of flight mass spectrometry at MAF/FM volumetric concentrations of 5%, 10%, 25% and 50%. Positive and negative polarities yield comparable f($q/m$) curves, though the negative mode yields ~30% larger currents. On shifting from the highest to the lowest liquid flow rates, at which a cone-jet is stable, the more conducting solutions evolve from mostly drop to primarily ion emission. A purely ionic regime is not reached under any condition, but the drops achieve unusually high $q/m$. As a result, these sprays have excellent electrical propulsion characteristics, some being able to cover a 25-fold range of average $q/m$ with a propulsive efficiency typically in the range of 80%. Results of formamide electrolytes with formates and nitrates of several other amines are more briefly reported.

## I. INTRODUCTION

Formamide is a highly polar solvent (dielectric constant $\varepsilon$=111) of low viscosity ($\mu$=3.76 cP at 20 °C), which dissolves large concentrations of many salts, reaching electrical conductivities $K$ well above 1 S/m. Although this conductivity does not match that of concentrated water electrolytes, formamide has much lower vapor pressure and can be exposed to a vacuum. It hence offers a rare opportunity to investigate the formation of nanojets and nanodrops from electrified liquid cones of low-viscosity and high-conductivity liquids in vacuum. Earlier studies with formamide electrolytes used salts made up of small ions, such as NaI[1] or ammonium acetate.[2] Under favorable conditions of high NaI concentration and low liquid flow rate, the current is dominated by ions, but does not present the purely ionic (PI) emissions typical of liquid metals.[3] This special regime must involve a *closed meniscus* topology (Figure 1a), with a steady tip that does not shed drops but is sharp enough to evaporate ions. This differs radically from the better studied *open meniscus* geometry of the cone-jet regime.[4,5] Here, the conical tip opens into a relatively long and initially steady jet, which becomes unstable and breaks into drops further downstream (Figure 1b). Over recent years, research on Taylor cones in vacuum has been dominated by electrical propulsion studies of room temperature molten salts displaying also the purely ionic regime[6,7,8,9,10,11,12] of liquid metals. The drop regime is nonetheless of considerable interest (as is its putative transition to the purely ionic regime), as it offers a controllable range of the charge over mass ratio $q/m$, while the purely ionic regime typically relies primarily on a single projectile of fixed $q/m$. Two main applications are currently driving interest on the drop regime. One relates to electrical propulsion in space, where the variable $q/m$ offers the possibility to control the specific impulse (thrust over mass flow rate ~ $(q/m)^{1/2}$) over a much wider range of values than in ion propulsion. The challenge, however, is to include within this controlable $q/m$ range values comparable to those typical of ion propulsion.[2] This ideal evidently requires extending the drop regime to very small sizes, approaching or even reaching the purely ionic regime.[13] The second application recently introduced by Gamero and colleagues is to use these nanodrops as hypervelocity projectiles for bombarding materials to either erode them or modify their surfaces.[14,15,16]

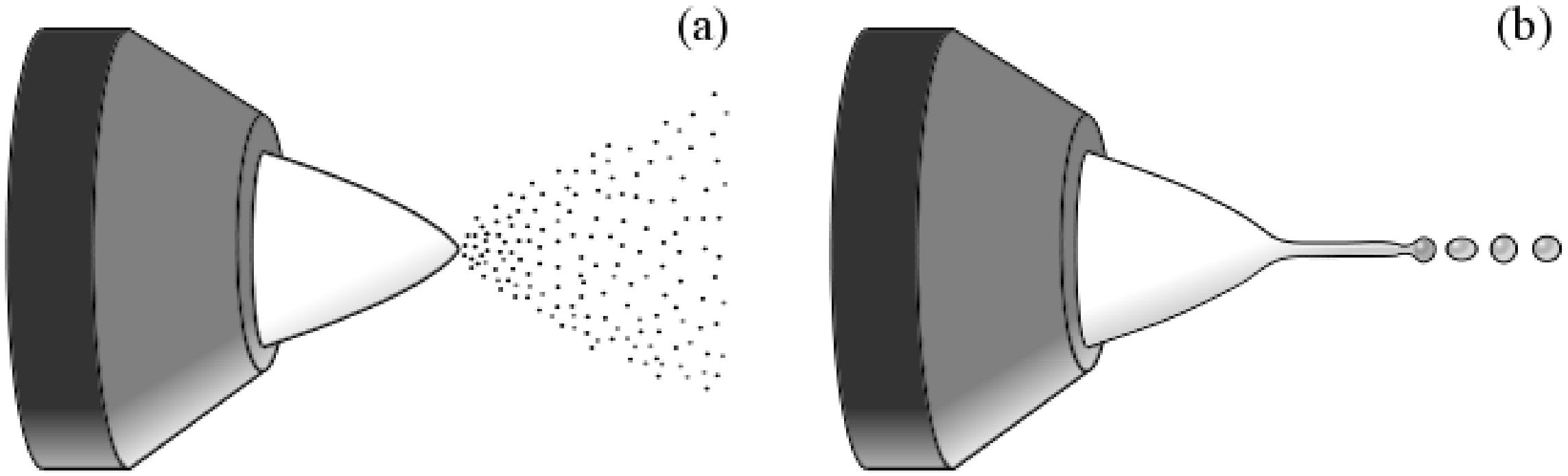

FIG. 1. Representation of the Taylor Cone supported in a sharpened needle tip. (a) *Closed meniscus* topology with ion evaporation from the tip of the cone. (b) *Open meniscus* topology with a jet that breaks into drops.

The present article is concerned with the charged particle beams produced by electrosprays of formamide combined with a variety of dissolved salts. We have previously given a summary account of some characteristics of such sprays when the salts were based on room temperature ionic liquids.[17,18] These studies were focused on materials capable of reaching the purely ionic regime, and included limited information relevant to the drop regime. In this regard, some of the most interesting formamide mixtures found involved methylammonium formate. Here we report in some detail their drop emission characteristics. Less singular results found with other salts combining several amines with formic acid will also be briefly described. It is important to note that all these salts belong to the class of so-called protic ionic liquids (PILs) as they are formed through proton transfer from an acid to a base.[19] Since the bases used are relatively weak, the reaction is partly reversible, leading to a finite equilibrium concentration of acid and base in the liquid.[20] Because both the acid and the base are volatile, unlike most salts, PILs have a finite volatility. For instance, as reported in Table 3 of Greaves et al. 2008,[19] the boiling point of methylammonium formate is about 182 °C, making it more volatile than formamide, see Table 1.

TABLE I. Liquid properties at room temperature (20ºC). Molecular weight $MW$, density $\rho$, viscosity $\mu$, conductivity $K$, surface tension $\gamma$, melting point $T_m$, boiling point $T_b$ and vapor pressure $p_v$.

| Liquid | $MW$ (Da) | $\rho$ (g/cm$^3$) | $\mu$ (cP) | $K$ (S/m) | $\gamma$ (mN/m) | $T_m$ (ºC) | $T_b$ (ºC) | $p_v$ (Pa) | Ref. |
|---|---|---|---|---|---|---|---|---|---|
| FM | 45.04 | 1.13 | 3.76 | - | 58.35 | 2.55 | 210.5 | 1.9 | 21 |
| MAF | 77.08 | 1.087[19], 1.12[a] | 17 | 4.38[19], 2.93[a] | 43.1 | 13 | 182 | - | 19 |

[a]Measured in the present study from the synthesized MAF.

## II. EXPERIMENTAL

All the experiments have been carried out in the system shown in Figure 2, similar to that of Gamero and Hruby[22]. The system is contained in a cylindrical vacuum chamber, with 25 cm ID (inner diameter) and 60 cm length. The electrospray experiment is made under vacuum, typically at a pressure under $10^{-5}$ torr. The conductive liquid is fed into the chamber from an external polypropylene vial through a silica capillary tube of 20 μm ID. The flow rate of pushed liquid is controlled by varying the pressure $\Delta P$ of gas in the polypropylene reservoir. The silica capillary tube inside the vacuum chamber passes through a straight metallic tube of 350 μm ID showing only the tip of silica at the extreme. This tip is sharpened like a needle and has a thin surface deposit of tin oxide which makes it conductive. Facing the silica needle perpendicularly to its axis is a grounded extractor electrode, while a high voltage $V_n$ is applied at the needle tip through the metallic tube. The high voltage source provides several kV in both positive and negative mode. The Taylor cone of the conductive liquid was supported on the tip of the silica needle inside the vacuum chamber. The charged electrosprayed particles entered into the larger vacuum chamber through a small hole on the extractor facing the tip of the needle and coaxial with it. Periodic interruption of the Taylor cone via a high voltage switch that grounded it permits analysis of the beam by time of flight mass spectrometry (TOF-MS). The periodic interruption has a frequency of 10 Hz, and the time to decrease the high voltage to ground is less than 50 ns. The emitted particles travel in free flight a longitudinal distance ($L$ = 12.3 cm) from the extractor into a collector electrode having a diameter of 21 cm. The current received by the collector electrode is represented as a function of the time $t$ following beam interruption. The trace $I(t)$ exhibits a flat region at small $t$, followed by a succession of steps, each associated to the different times of flight of the various particles forming the beam. Short flight times reveal the presence of ions, and long flight times that of drops. The collector was grounded through an electrometer, and was preceded shortly upstream by a grid held at a slightly negative voltage to avoid secondary electron emission. That grid is 65 % transparent, which is taken into account to correct the measurements. In addition, the ion current intercepted by the extractor directly has been measured and controlled during all the experiments, and is less than 2 nA.

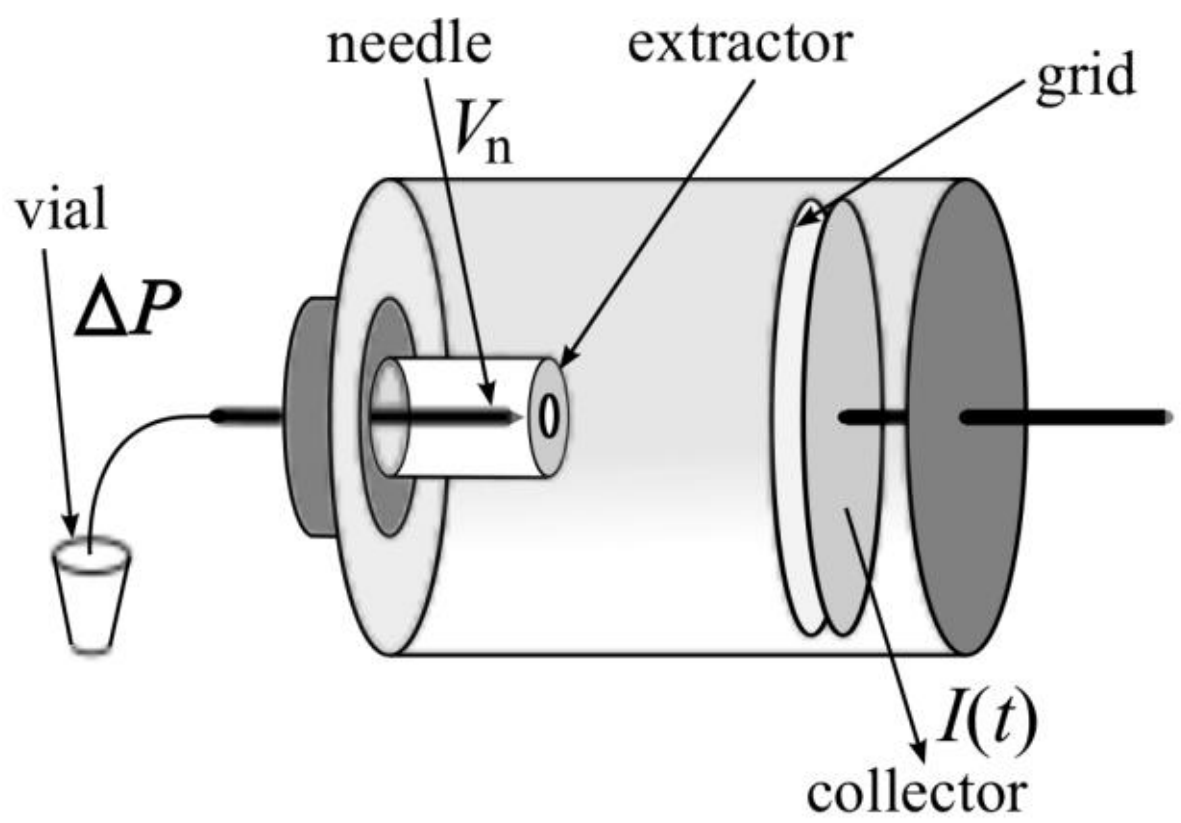

FIG. 2. Experimental setup for the time of flight mass spectrometer (TOF-MS)

The liquid flow rate and the various propulsive parameters of interest have been inferred by suitable integration of the $I(t)$ curves under the assumption that the energy of the charged drops is equal to their charge $q$ times $V_a$, which is the needle voltage $V_n$ minus a small voltage drop for beam formation which we take to be 200 V.[22] In addition, the $I(t)$ curves have been multiplied by a factor of 1.538 to correct for the 65% lost current into the grid.

The conductive liquids primarily studied in this work are mixtures of formamide (FM) with methylammonium formate (MAF). Pure FM is a non-conductive polar solvent while MAF is a protic ionic liquid with high conductivity, see Table I. Both are colorless liquids at room temperature and completely miscible with each other. MAF has been synthesized in our laboratory[17] from methylamine and formic acid that were purchased from Aldrich and Fluka. The synthesis is based on the neutralization of an amine with the acid, which requires certain essential safety precautions described by Angell and colleagues.[20] Note in particular that the reaction is highly exothermic and must proceed at low temperatures and small volumes. The amine and acid are volatile and toxic and must be handled under a fume hood. Because of its high volatility the amine was purchased in aqueous solution, and the final product had to be dried under vacuum for several hours. The density and conductivity were measured, and are compared in Table I with published data.[19] Four mixtures of FM with MAF have been studied in order to obtain liquids of variable conductivity (Table II).

TABLE II. Measured conductivities for mixtures MAF/FM versus volume % of MAF.

| MAF/FM | 5% | 10% | 25% | 50% | Pure MAF |
|---|---|---|---|---|---|
| $K$ (S/m) | 0.97 | 1.52 | 2.43 | 2.80 | 4.38 |

## III. RESULT

Figure 3 includes all measurements TOF-MS carried with MAF/FM mixtures, with volumetric concentration of salt, given in each figure, increasing from top to bottom. The Figures on the left are for positive sprays and those on the right for negative sprays. The two polarities show a comparable behavior, though with a current typically 30% higher in negative mode. Focusing first on the two top figures, for 5% MAF/FM, we read the total current on the left ($I_0$), which identifies the experiment in Table IV, and helps determine its other characteristics, particularly the flow rate of injected liquid. All curves exhibit a small step at about 5 μs. Its height is the ion current, $I_{ion}$, about 20 nA in the two top figures, with almost no dependence on liquid mass flow rate $m' = \rho Q$. The tall and broad steps seen at times $t$ of tens of μs correspond to the flight times of charged drops. They show the expected reduction in mean flight time $t^*$ with decreasing liquid flow rate $Q$. Within experimental error, the total current and $t^*$ decrease monotonically with decreasing $Q$. The curves for 10% MAF/FM show a behavior similar to that for 5% salt, though with the richer $I_{ion}(Q)$ dependence shown in Figure 4. At high $Q$, $I_{ion}$ takes an asymptotic value $I_{ion}^{\infty}$, which, as before, is relatively independent of $Q$. Now, however, below a critical flow rate $Q^*$, the $I(t)$ curves exhibit a rapidly increasing ion current with decreasing $Q$. The behavior at even higher salt concentrations follows a similar pattern, with a certain dependence on MAF concentration (or conductivity $K$) of the characteristic quantities $I_{ion}^{\infty}$, $t^*$ and $Q^*$: $I_{ion}^{\infty}$ increases, the drop flight times $t^*$ decrease, and there is a widening of the range of flow rates $Q < Q^*$ where ion evaporation from the meniscus takes over. These trends are similar to those previously observed in formamide-NaI solutions.[1] The more or less constant ion current at high flow rates has been attributed to emissions not from the high field region at the meniscus neck, but rather from either the drops immediately after jet breakup, or from the singular breakup region where the drop detaches from the jet. The sharper increase in ion current at decreasing liquid flow rate has been explained as due to ion evaporation directly from the meniscus neck, where the local electric field presents a maximum $E_{max}$, which increases as the jet becomes narrower with diminishing $Q$ or increasing $K$. Gamero et al.[1] have argued that the maximum local field at the meniscus neck is proportional to the measurable quantity $E_K$:

$$E_{\max} = \varphi(\varepsilon) E_K \; ; \; E_K = \frac{\gamma^{1/2} K^{1/6}}{\varepsilon_0^{2/3} Q^{1/6}} , \qquad (1, 2)$$

a point approximately confirmed in calculations reported by Guerrero et al.,[13] who also reports the value of the proportionality parameter for propylene carbonate electrolytes: $\varphi(65) \sim 0.76$. Ion evaporation directly from the meniscus neck is expected when $E_K$ reaches values over 1 V/nm. This can be observed in Figure 5 where $I_{ion}$-$I_{ion}^{\infty}$ increases quickly for $E_K > 1$ V/nm. The $E_K$ values reported for each experiment in Tables IV show that $E_K$ values are over 1 V/nm at low flow rates in 10% and 25% MAF/FM, while in the case of 50% MAF/FM the value of $E_K$ is higher than 1 V/nm for all flow rates. The anomalous increase of $I_{ion}$ at the smallest value of $E_K$ is associated to the highest liquid flow rates used, at which the mechanism of jet breakup becomes more disordered. This ion current is therefore probably not coming from the jet neck, but has an origin similar to that of $I_{ion}^{\infty}$.

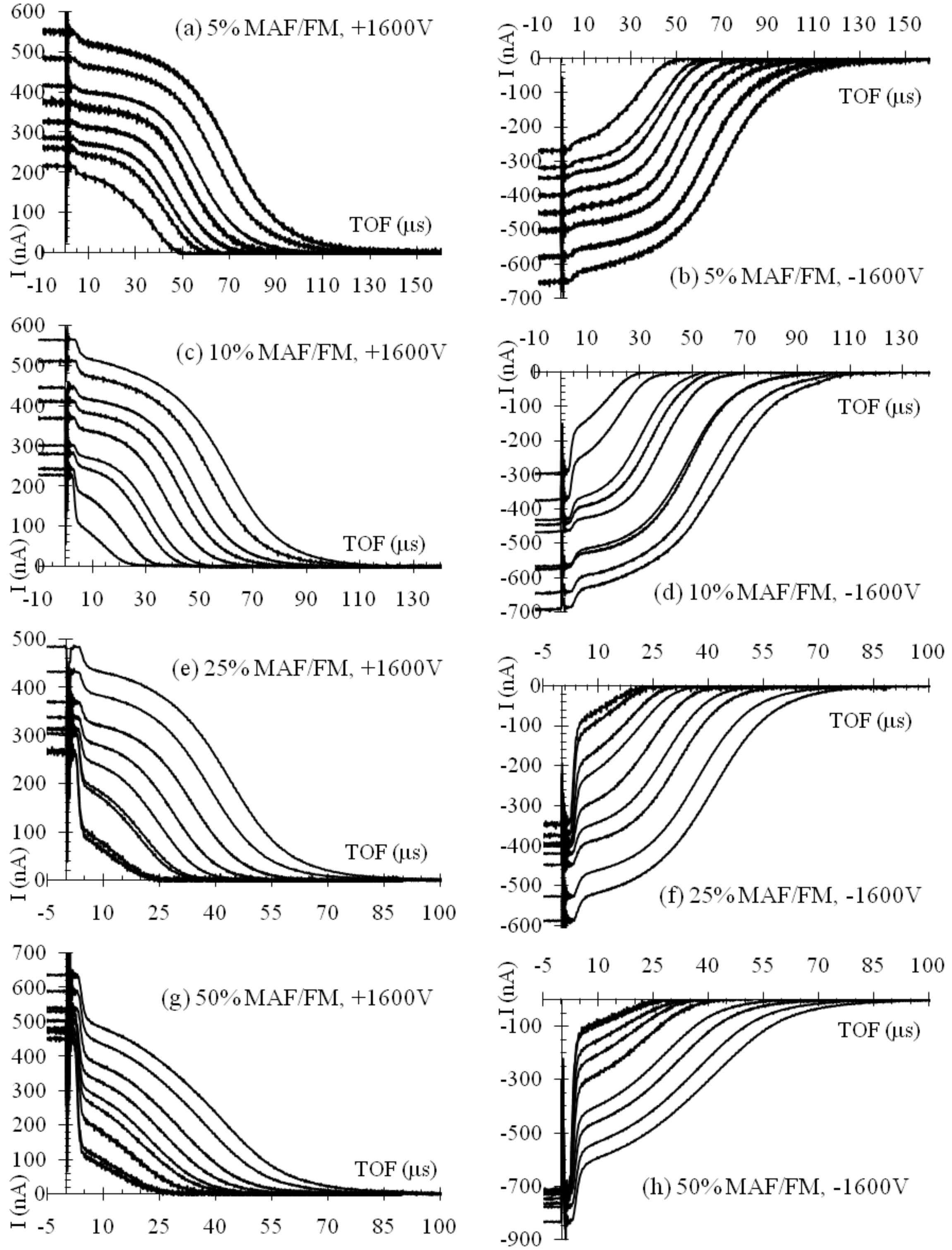

FIG. 3: Time of flight (TOF) distributions of the charged particle beams produced by mixtures of formamide with methylammonium formate at the various concentrations indicated, in positive (left) or negative polarity (right). Each of the curves shown for given liquid composition corresponds to different flow rates of liquid flowing through the cone-jet. Each is characterized in Table IV by the total beam current, that can be read on the left end of the curves.

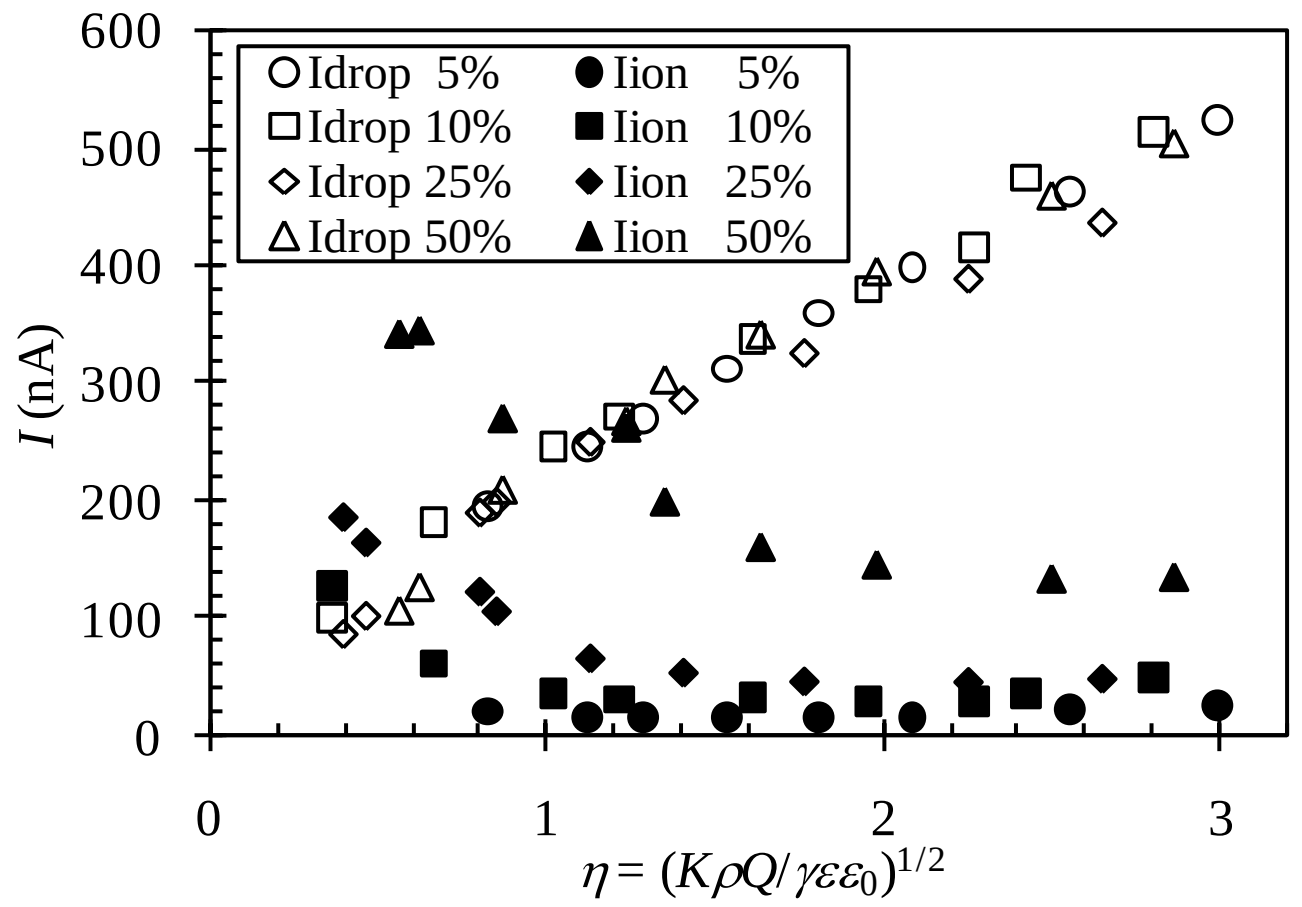

FIG. 4. Drop current ($I_{drop}$) and ion current ($I_{ion}$) versus the dimensionless flow rate parameter $\eta = (\rho KQ/\gamma\varepsilon\varepsilon_o)^{1/2}$ for MAF/FM mixtures in positive mode.

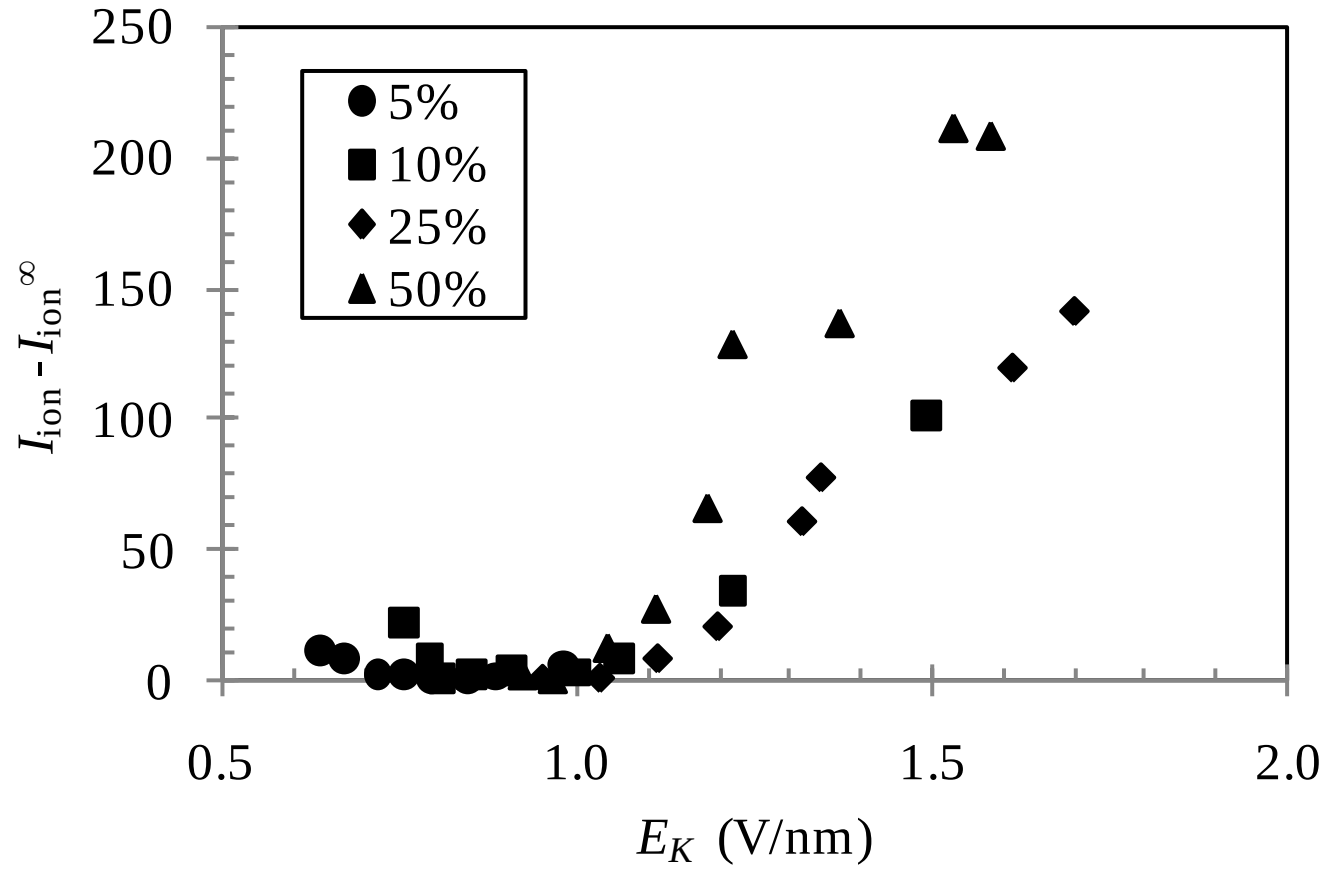

FIG. 5. Difference between the ion current ($I_{ion}$) and its asymptotic value ($I_{ion}^{\infty}$) versus $E_K$ (Equation 2) for MAF/FM mixtures in positive mode.

Figure 4 shows that the drop current ($I_{drop}$) is approximately linear in the dimensionless flow rate parameter $\eta = (\rho KQ/\gamma\varepsilon\varepsilon_o)^{1/2}$, very much as in the better studied limit when no ion current is produced at all.[23,24,25] There are nonetheless some noteworthy variations. The relatively ion-free solution with 5% MAF ($K$ = 0.97 S/m) extends its operating range only down to $\eta_{min}$ ~ 0.8, slightly above the value $\eta_{min}$ ~ 0.6 previously reported for formamide at lower conductivities.[23] The solutions with 10% and 25% MAF yielding a fair ion current widen that range substantially, down to $\eta_{min}$ ~ 0.4. This widening goes in the opposite direction of conventional electrosprays (with no ion emission), where an increase in $K$ increases $\eta_{min}$.[23] The difference is undoubtedly due to the ejection of ions, which changes the boundary condition for the charge transport problem from the bulk of the conducting liquid towards its interface with the vacuum. The advantage of a decreased minimum flow is partly lost at the highest conductivity. This may be due to the increase in viscosity of the 50% mixture with respect to pure formamide, which is known to increase $\eta_{min}$ in conventional electrosprays. There are other subtle effects of $K$ in Figure 4, such a slight reduction and a slight increase of the high $\eta$ current for 25% and the 10% solutions, and a depression of the low $\eta$ current for the 50% solution. Notwithstanding this, to a first approximation, one may say that the behavior for $\eta \geq 1$ is very much as in the absence of ion evaporation. While the drop current decreases with diminishing $Q$, the ion current $I_{\text{ion}}$ increases, reaching values above $I_{\text{drop}}$ at sufficiently low liquid flow rates and high conductivities. In this later cases, the f($q/m$) of the electrospray is dominated by ion current reaching values higher than pure colloidal electrosprays.

Figure 6 shows two important propulsive parameters of the mixtures studied, the specific impulse ($I_{\text{sp}}$) and the propulsion efficiency ($\eta_p$), as function of the thrust ($T$).The mass flow rate ($m'$) and $T$ are computed from $I(t)$ curves according to equations[22]

$$m' = \frac{4V_a}{L^2}\int_0^\infty tI(t)\mathrm{d}t \;;\; T = \frac{2V_a}{L}\int_0^\infty I(t)\mathrm{d}t\,, \qquad (3, 4)$$

whereas $I_{\text{sp}}$ and $\eta_p$ are defined with

$$I_{\text{sp}} = \frac{T}{\mathrm{g}m'}\;;\; \eta_p = \frac{T^2}{2m'\mathrm{V}_a I_0}\,, \qquad (5, 6)$$

where g = 9.8 m/s$^2$ is the gravitational acceleration. The propulsive parameters are also included in Table IV. For the 5% and 10% MAF/FM mixtures, one single Taylor cone has a wide range of thrust between 0.15 to 0.82 μN with a high propulsion efficiency of 80% and specific impulses that can reach values over 400 s at low flow rates. The 50% MAF/FM mixture has lower efficiency at all flow rates, due to the high ion evaporation. The best performance is achieved at 25% MAF/FM, with high efficiency 80% at the lowest specific impulses (from 200 s to 400 s), and a maximum $I_{\text{sp}}$ 1100 s with an efficiency of 50% at low flow rates (low thrust). In negative mode, the propulsive parameters have almost the same ranges as the positive ones. The maximum $T$ is 1 μN (at 5% concentration) and the maximum $I_{\text{sp}}$ is 1345 s (at 50% concentration).

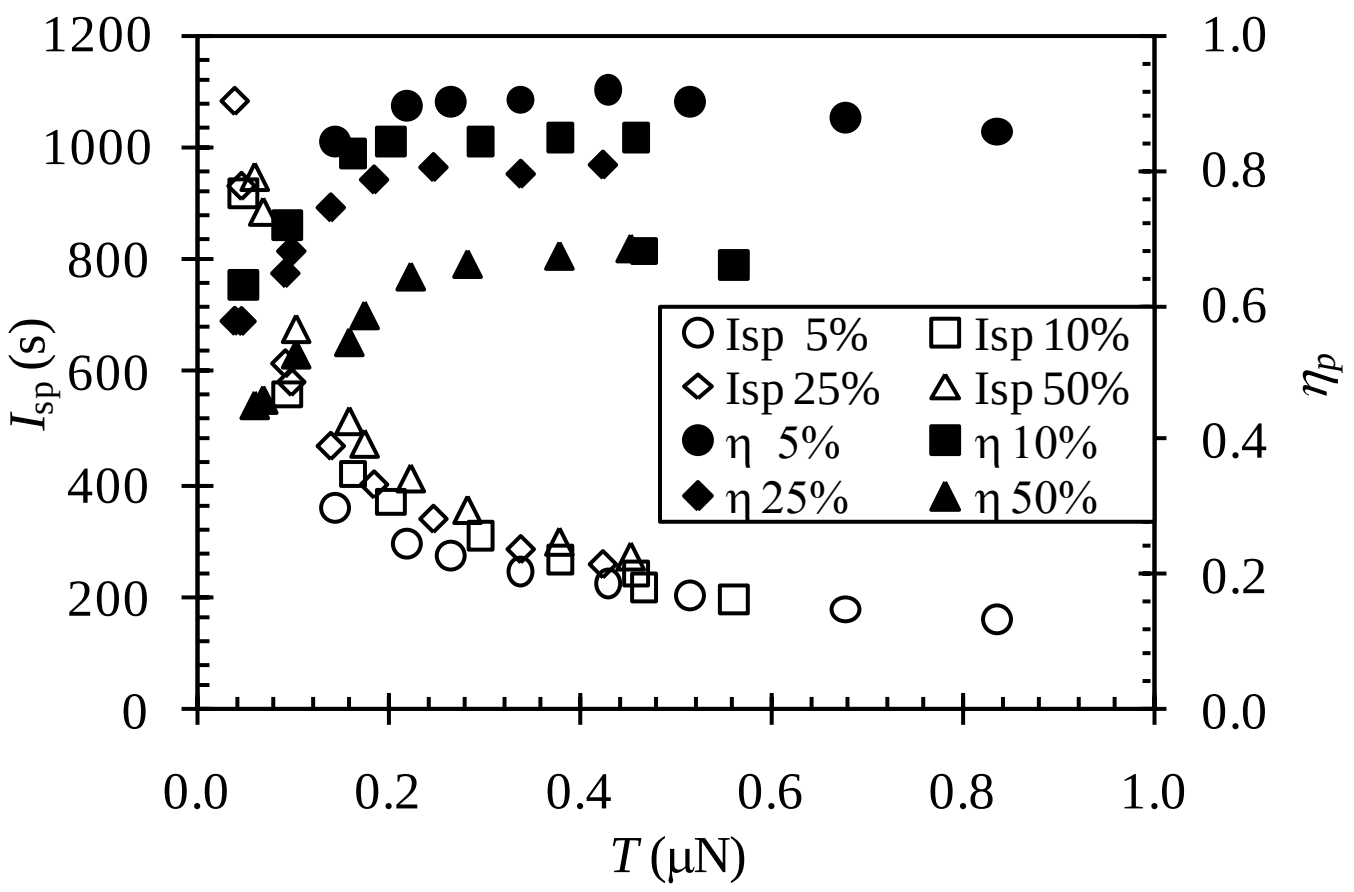

FIG. 6. Positive-mode propulsive properties of our four MAF/FM mixtures.

Several other formamide mixtures with formates and nitrates have been similarly studied, with key propulsive parameters reported in Figure 7. The thrust range achieved is between 0.1 to 0.9 μN per Taylor-cone. Figure 7 shows the $\eta_p$ versus $I_{sp}$ for the most promising mixtures achieving the highest $I_{sp}$. Among formates, the 25% MAF/FM is the most interesting with $I_{sp}$ reaching up to 1100 s, and high $\eta_p$ at large thrust (low $I_{sp}$). Dimethylammonium formate DMAF, ethylammonium formate EAF and ammonium formate AF maintain an efficiency over 80% in an interesting $I_{sp}$ range from 150 s to 500 s. Therefore, these mixtures can be suitable for micro-thrust systems that require the mentioned specific range of $I_{sp}$. Mixtures of nitrates with formamide have a similar behavior as formates. It is noteworthy that the mixture of ammonium nitrate with formamide (15:100 AN/FM) has a range of $I_{sp}$ from 150 s to 750 s, but for large and low thrust (low and large $I_{sp}$ respectively) the efficiency falls below 80%. The excellent behavior of AN/FM mixtures (4.23 M ~ 30:100 weight) has been previously reported by Bocanegra et al.,[2] who attributed it to the almost complete absence of ions. Our measurements, however do show an important ion current when the concentration of AN increases (increasing conductivity $K$). Although the reason for this discrepancy is unclear, the earlier results may correspond to an unusual regime, as $\eta_p$ had the uncommon feature of increasing with increasing $I_{sp}$. All other liquis studied here as well as by Bocanegra et al.[2] show a decreasing $\eta_p$ with increasing $I_{sp}$. Figure 8 shows the TOF-MS spectra of two mixtures of AN/FM at two concentrations. For 5:100 AN/FM (Figure 8a), the ion current is almost negligible and the large $t^*$ reveals a small $I_{sp}$ below 250 s. On the other hand, for 15:100 AN/FM (Figure 8b), the ion current increases at decreasing flow rate due to ion evaporation from the meniscus neck. The anomalous ion current at large flow rate is associated to a disordered mechanism of jet breakup.

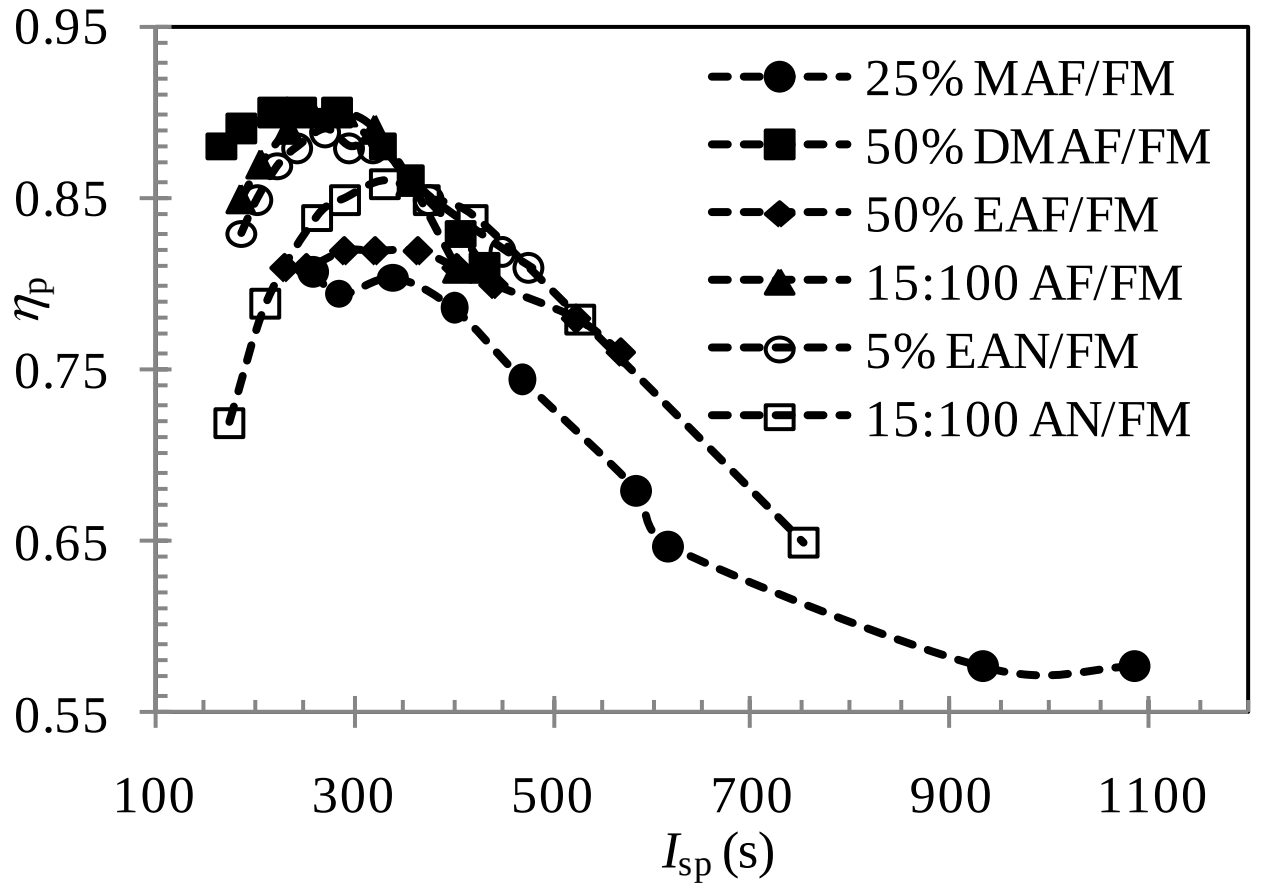

FIG. 7 Propulsion properties of different mixtures of formamide with formates and nitrates. MAF = methylammonium formate, DMAF = Dimethylammonium formate, EAF = ethylammonium formate, AF = ammonium formate, EAN = ethylammonium nitrate, AN = ammonium nitrate. Mixtures of AF/FM and AN/FM are indicated with the percent per weight 15:100.

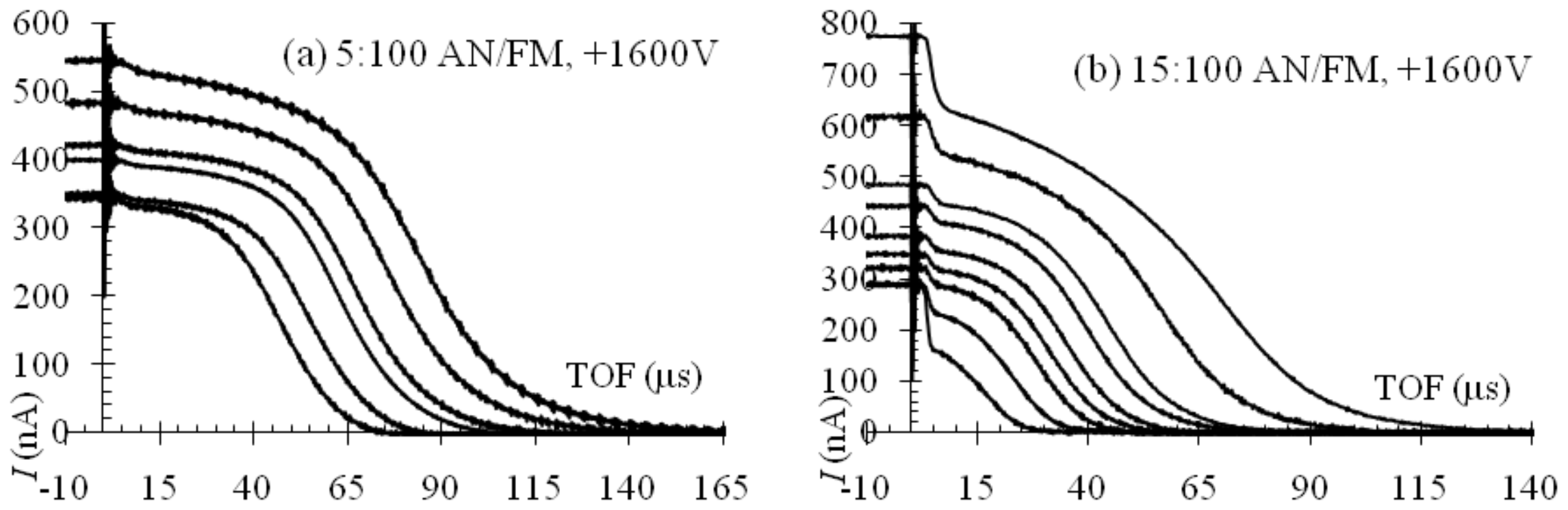

FIG. 8. Positive mode time of flight traces for two mixtures of formamide with ammonium nitrate.
(a) 5:100 AN/FM and (b) 15:100 AN/FM

## IV. DISCUSSION

Emission of ion current is observed in all the TOF-MS spectra of MAF/FM mixtures studied. At high flow rate and low conductivity the total current follows the scaling law of pure colloidal electrospray. Therefore, the small ion evaporation current observed must occur after the jet breaks into drops. In this case, the maximum electric field located at the surface of a drop is $E_d = q_d/4\pi\varepsilon_0 r_d^2$, where $q_d$ is the drop charge and $r_d$ is the drop radius estimated with the scaling law $r_d = 0.5g(\varepsilon)(Q\varepsilon\varepsilon_0/K)^{1/3}$, $g(111) = 0.883$ for formamide.[23,24,25] Ion evaporation from drops is evidently limited to the small fraction of the total drop charge whose loss suffices to decrease the surface electric field and therefore quench further charge loss. We follow Gamero et al.[1] in interpreting the measured values $I_{\text{ion}}^{\infty}$ at various flow rates and mixture compositions in order to infer the activation energy ($\Delta G$) of the evaporated ions from the drops. Following Iribarne and Thomson,[26] the charge balance for a drop is governed by

$$\frac{\mathrm{d}q_d}{\mathrm{d}t} = \frac{q_d k_B T_0}{h} \exp\left(-\frac{\Delta G - G_E\left(E_q\right)}{k_B T_0}\right), \text{ with } G_E\left(E_q\right) = \left(\frac{e^3 E_q}{4\pi\varepsilon_0}\right)^{1/2}, \qquad (7, 8)$$

where $k_B$ and $h$ are the Boltzmann and Planck constants, $T_0$ is the absolute temperature, and $e$ is the elementary charge. The solution with the initial condition $q_d = q_{d0} = 4\pi r_d^3 I_0/3Q$ can be simplified in the following expression

$$\frac{\exp\left(\psi_{d0} I_{\text{ion}}/2I_0\right)}{1-\left(I_{\text{ion}}/2I_0\right)} - 1 = \mathrm{n}\,\psi_{d0} \exp\left(\psi_{d0}\right), \qquad (9)$$

where $\psi_{d0} = e^{3/2} q_{d0}^{1/2}/4\pi\varepsilon_0 r_d k_B T_0$ and $\mathrm{n} = \exp(-\Delta G/k_B T_0)\, t_f\, k_B T_0/2h$ is a constant that depends on the activation energy $\Delta G$ and evaporation time $t_f$. The value of n can be calculated using the experimental values $Q$ , $I_0$ and $I_{\text{ion}}$ of the MAF/FM mixtures. Therefore, one can estimate the activation energy $\Delta G$ when the evaporation time $t_f$ is taken to be 1 ns.[1] This latter choice has little influence on the final value of $\Delta G$, as a factor of 10 in $t_f$ shifts $\Delta G$ by only 0.06 eV. The values of $\Delta G$ are reported in Table IV for the experiments in the regime of ions evaporated from drops, $\eta > 1$ and $E_K < 1$ V/nm. The mean value of $\Delta G$ is 1.98 eV with a standard deviation of 0.14 eV.

The smaller mass $m_{\text{ion}}$ of the ions evaporated from the drops can be estimated from the time $t_{\text{ion}}$ of the first step in TOF curves, $m_{\text{ion}} = 2qV_a(t_{\text{ion}}/L)^2$. This mass is reported in Table III (in atomic mass units; $MW_{\text{ion}}$), showing how it decreases upon increasing the percentage volume of MAF (increasing conductivity). The smaller $t_{\text{ion}} = 1.35$ is obtained from the spectra of 50% MAF/FM, which leads to a $MW_{\text{ion}}$ close to the cation MA+ = 32.03 Da. For mixtures with lower MAF concentrations, the evaporated ions have a mass higher than the cation, probably due to attachment of several formamide molecules, and perhaps also one or several neutral ion pairs of MAF.

TABLE III. Time of ion step and calculated mass weight.

| MAF/FM | $t_{\text{ion}}$ (μs) | $MW_{\text{ion}}$ (Da) |
|---|---|---|
| 5% | 3.95 | 276.6 |
| 10% | 2.5 | 110.8 |
| 25% | 2.2 | 85.8 |
| 50% | 1.35 | 32.55 |

At low flow rates and in high conductivity mixtures, the ion current increases quickly with decreasing $Q$, and the total current exceeds substantially the value for the usual cone-jet regime, see Figure 4. In this situation the ions must evaporate from somewhere before the jet break into drops. The maximum electric field is located at the meniscus neck and can be evaluated with equation 2. The current carried by the jet decreases when ions evaporate from the meniscus neck, therefore the evaporated ions from drops also decrease. Here we have not distinguished between the ionic current originating from either the drops or the meniscus neck, though it would be safe to assign to the neck at least all the ion current minus $I_{\text{ion}}^{\infty}$.

All the mixtures studied of MAF/FM can be sprayed with comparable stability in positive and negative mode. The propulsive parameters inferred from the TOF curves of Figure 3 are similar in negative and in positive mode. This feature is important because a propulsive system used in a satellite in bipolar mode (alternating positive and negative mode) would not accumulate positive or negative charge.

Finally, we have estimated the mass evaporated from the meniscus as a result of the finite volatility of formamide and MAF. Mass can evaporated from drops or from the cone-jet surface. The rate of change in drop radius as a result of mass evaporation can be calculated as $r_d' = -m_{\text{FM}}^{1/2} p_v(T_0)/\rho(2\pi k_B T_0)^{1/2}$, where $m_{\text{FM}}$ is the molecular mass of formamide.[27] Using the physical properties of formamide at room temperature, see Table I, and $t^*$ as drop flight time, the drop radius decreases by 0.1 nm. Therefore, the mass evaporated from drops is negligible, below 2% of the initial mass. On the other hand, the mass evaporated from the cone-jet is expected to be almost constant, because the same needle of 20 μm ID was used in all the experiments and thus the cone-jet area should be mostly the same. The mass evaporated from the cone-jet can be determined from the difference between the mass injected into the meniscus and that received by the collector, which we extract from the TOF curves. The flow rate $m'$ calculated in Table IV varies linearly with the pressure applied to the vial $\Delta P$, as predicted by Poiseuille's law, see Figure 9. The trend lines for each mixture cut the $y$ axis below the origin, at values from 0.05 to 0.025 μg/s. Assuming that evaporated mass through the cone-jet surface is constant with $m'$, these values can be directly related with the evaporated mass. Evaporated mass from the cone-jet surface is similar than the reported by Gamero and Hruby[22] of 0.04 μg/s with formamide mixtures in needles of 22 μm ID. Although the evaporated mass is below 10% of the injected mass at high flow rates, it becomes relevant (85%) at low flow rates of highly conductive mixtures, decreasing the propulsive parameters such as $I_{sp}$ and $\eta_p$. In order to maintain the good propulsive parameters potentially offered by MAF/FM mixtures, the propulsion systems must be designed with needle tip diameters several times smaller than 20 μm.

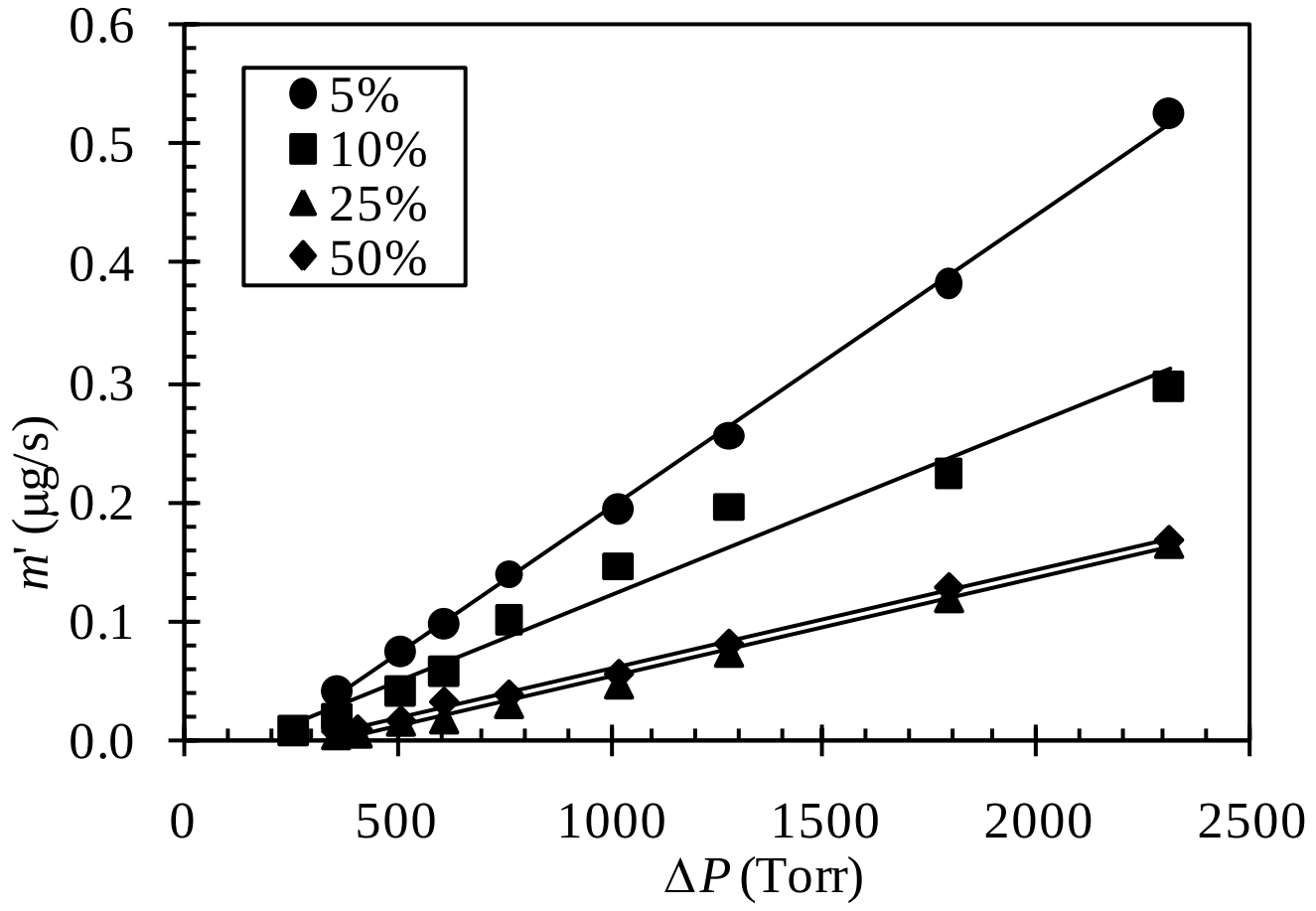

FIG. 9. Mass flow rate ($m'$) as function of pressure applied in the vial ($\Delta P$). Linear trend lines are added to each serial data.

## V. CONCLUSIONS

A regime from Taylor cones in vacuum with ion evaporation has been studied for formamide mixtures with methylammonium formate at different concentrations. The ion current has been measured directly from TOF-MS spectra, and two ion evaporation regimes have been detected. In low concentration (low conductivity) mixtures and at large flow rates, ions evaporate from drops with a constant ion current. At higher concentration (higher conductivity) and low flow rates, ions evaporate from the meniscus and the ion current increases quickly when the flow rate decreases. This latter regime occurs when $E_K$ > 1 V/nm, and the scaling law for emitted current is not valid. An activation energy $\Delta G$ of 1.98 eV has been estimated form the ion current when ions evaporate from drops.

Formamide mixtures have been characterized via propulsion properties inferred from TOF-MS curves in positive and negative mode. The mixture with 25% MAF/FM achieves exceptional propulsive properties in a wide range of conditions. The high specific impulse with moderate efficiency makes this mixture a good candidate as propellant fuel in spacecraft. In addition, this propellant fuel can work in positive and negative mode to avoid charging the spacecraft. Others mixtures of formamide with formates and nitrates are reported briefly.

Finally, although the evaporated mass due to the finite volatility of the mixtures can be a problem for propulsion systems working at low flow rates, this limitation can in principle be moderated by reducing the size of the needle tip.

**ACKNOWLEDGMENTS**

This work has been partly supported by U.S. AFOSR contract FA9550-09-C-0178 to Alameda Applied Sciences, through a subcontract to Yale, and by the Spanish Ministry of Science via DPI2010-20450-C03-01.

TABLE IV. Currents and propulsion characteristics of MAF/FM mixtures in positive mode (+1600 V) with a needle 20 μm ID and $L$ = 12.37 cm.

| $\Delta P$ (Torr) | $I_0$ (nA) | $I_{drop}$ (nA) | $I_{ion}$ (nA) | $t^*$ (ms) | $m'$ (μg/s) | $T$ (μN) | $I_{sp}$ (s) | $\eta_p$ (%) | $(q/m)_{drop}$ (C/kg) | $\eta$ | $E_K$ (V/nm) | $\Delta G$ (eV) |
|---|---|---|---|---|---|---|---|---|---|---|---|---|
| 5% MAF/FM ($K$ = 0.97 S/m) | | | | | | | | | | | | |
| 2311 | 551.1 | 524.5 | 26.7 | 69.7 | 0.528 | 0.838 | 159 | 85.7 | 1125 | 2.99 | 0.637 | 1.70 |
| 1794 | 486.8 | 463.7 | 23.1 | 64.0 | 0.383 | 0.678 | 177 | 87.9 | 1334 | 2.55 | 0.672 | 1.76 |
| 1277 | 415.8 | 398.8 | 17.0 | 57.1 | 0.256 | 0.516 | 202 | 90.1 | 1676 | 2.08 | 0.719 | 1.85 |
| 1019 | 377.0 | 359.9 | 17.1 | 52.7 | 0.193 | 0.429 | 222 | 91.8 | 1968 | 1.81 | 0.754 | 1.92 |
| 760 | 328.4 | 313.0 | 15.4 | 47.7 | 0.139 | 0.338 | 243 | 90.4 | 2402 | 1.53 | 0.796 | 1.99 |
| 608 | 286.8 | 271.1 | 15.7 | 43.4 | 0.097 | 0.264 | 272 | 90.1 | 2901 | 1.28 | 0.845 | 2.07 |
| 506 | 264.3 | 247.6 | 16.7 | 40.6 | 0.074 | 0.219 | 296 | 89.3 | 3315 | 1.12 | 0.884 | 2.15 |
| 354 | 216.2 | 195.3 | 20.9 | 33.2 | 0.040 | 0.143 | 358 | 84.1 | 4958 | 0.82 | 0.980 | - |
| 10% MAF/FM ($K$ = 1.52 S/m) | | | | | | | | | | | | |
| 2311 | 566.0 | 516.8 | 49.2 | 59.3 | 0.2974 | 0.562 | 193 | 65.9 | 1554 | 2.81 | 0.756 | 1.89 |
| 1794 | 511.9 | 476.8 | 35.2 | 53.5 | 0.2229 | 0.467 | 214 | 67.8 | 1909 | 2.43 | 0.793 | 1.96 |
| 1277 | 445.7 | 418.3 | 27.4 | 47.9 | 0.1949 | 0.458 | 240 | 85.1 | 2382 | 2.27 | 0.811 | 1.91 |
| 1019 | 410.5 | 381.4 | 29.1 | 43.0 | 0.1449 | 0.379 | 267 | 85.1 | 2956 | 1.96 | 0.852 | 2.00 |
| 760 | 369.2 | 338.4 | 30.9 | 37.8 | 0.0984 | 0.295 | 306 | 84.4 | 3825 | 1.62 | 0.909 | 2.13 |
| 608 | 302.1 | 272.2 | 29.8 | 31.4 | 0.0556 | 0.200 | 367 | 84.3 | 5543 | 1.21 | 0.999 | 2.29 |
| 506 | 281.2 | 246.1 | 35.1 | 28.3 | 0.0393 | 0.161 | 417 | 82.5 | 6824 | 1.02 | 1.059 | - |
| 354 | 242.5 | 181.5 | 61.0 | 20.8 | 0.0167 | 0.092 | 560 | 71.9 | 12631 | 0.67 | 1.221 | - |
| 252 | 228.9 | 100.0 | 128.9 | 15.0 | 0.0050 | 0.045 | 921 | 62.7 | 24288 | 0.36 | 1.495 | - |
| 25% MAF/FM ($K$ = 2.43 S/m) | | | | | | | | | | | | |
| 2311 | 483.9 | 436.8 | 47.2 | 42.8 | 0.1652 | 0.426 | 258 | 80.8 | 2983 | 2.65 | 0.901 | 1.91 |
| 1794 | 433.6 | 389.1 | 44.5 | 38.0 | 0.1193 | 0.340 | 285 | 79.5 | 3785 | 2.25 | 0.952 | 1.99 |
| 1277 | 370.9 | 326.1 | 44.8 | 33.2 | 0.0733 | 0.248 | 339 | 80.5 | 4958 | 1.76 | 1.032 | - |
| 1019 | 338.4 | 286.1 | 52.3 | 28.1 | 0.0465 | 0.186 | 400 | 78.6 | 6921 | 1.40 | 1.114 | - |
| 760 | 315.3 | 250.7 | 64.6 | 23.9 | 0.0300 | 0.141 | 469 | 74.5 | 9567 | 1.13 | 1.198 | - |
| 608 | 304.7 | 199.9 | 104.8 | 18.6 | 0.0170 | 0.099 | 583 | 68.1 | 15796 | 0.85 | 1.317 | - |
| 506 | 312.3 | 190.7 | 121.6 | 19.1 | 0.0151 | 0.093 | 616 | 64.8 | 14980 | 0.80 | 1.343 | - |
| 404 | 266.6 | 103.0 | 163.6 | 14.3 | 0.0050 | 0.047 | 933 | 57.7 | 26724 | 0.46 | 1.613 | - |
| 354 | 272.9 | 87.7 | 185.2 | 13.9 | 0.0037 | 0.040 | 1084 | 57.8 | 28285 | 0.39 | 1.700 | - |
| 50% MAF/FM ($K$ = 2.80 S/m) | | | | | | | | | | | | |
| 2311 | 640.3 | 504.5 | 135.9 | 38.3 | 0.1671 | 0.452 | 271 | 68.5 | 3725 | 2.86 | 0.921 | 1.99 |
| 1794 | 594.3 | 459.9 | 134.4 | 34.7 | 0.1274 | 0.378 | 297 | 67.4 | 4539 | 2.49 | 0.964 | 2.07 |
| 1277 | 541.7 | 395.3 | 146.4 | 29.8 | 0.0800 | 0.282 | 353 | 66.1 | 6154 | 1.98 | 1.042 | - |
| 1019 | 502.9 | 341.4 | 161.5 | 26.7 | 0.0546 | 0.223 | 409 | 64.3 | 7666 | 1.63 | 1.110 | - |
| 760 | 503.2 | 303.0 | 200.3 | 23.3 | 0.0372 | 0.175 | 471 | 58.5 | 10066 | 1.35 | 1.183 | - |
| 608 | 531.0 | 267.6 | 263.3 | 21.5 | 0.0312 | 0.159 | 511 | 54.4 | 11822 | 1.23 | 1.219 | - |
| 506 | 480.6 | 209.2 | 271.4 | 18.6 | 0.0154 | 0.104 | 675 | 52.6 | 15796 | 0.87 | 1.370 | - |
| 404 | 472.6 | 126.1 | 346.4 | 14.1 | 0.0079 | 0.070 | 883 | 45.8 | 27488 | 0.62 | 1.532 | - |
| 354 | 449.6 | 106.1 | 343.5 | 14 | 0.0064 | 0.061 | 945 | 45.0 | 27882 | 0.56 | 1.585 | - |